# Journal of Optics
# Special issue on High Resolution Optical Imaging

## Polarized light imaging of birefringence and diattenuation at high resolution and high sensitivity

Shalin B. Mehta[1], Michael Shribak[1], and Rudolf Oldenbourg[1,2]
1 Marine Biological Laboratory, Woods Hole MA 02543, and
2 Physics Department, Brown University, Providence RI 02912

Corresponding Author:
Rudolf Oldenbourg
Marine Biological Laboratory
7 MBL Street
Woods Hole MA 02543, USA

Email: rudolfo@mbl.edu
Phone: 1-508-289-7426
Fax: 1-508-540-6902

### Abstract

Polarized light microscopy provides unique opportunities for analyzing the molecular order in man-made and natural materials, including biological structures inside living cells, tissues, and whole organisms. 20 years ago, the LC-PolScope was introduced as a modern version of the traditional polarizing microscope enhanced by liquid crystal devices for the control of polarization, and by electronic imaging and digital image processing for fast and comprehensive image acquisition and analysis. The LC-PolScope is commonly used for birefringence imaging, analyzing the spatial and temporal variations of the differential phase delay in ordered and transparent materials. Here we describe an alternative use of the LC-PolScope for imaging the polarization dependent transmittance of dichroic materials. We explain the minor changes needed to convert the instrument between the two imaging modes, discuss the relationship between the quantities measured with either instrument, and touch on the physical connection between refractive index, birefringence, transmittance, diattenuation, and dichroism.

## 1. Introduction

The polarized light microscope is used to analyze the anisotropy of a specimen's optical properties such as refraction and absorption. Optical anisotropy is a consequence of molecular order, which can render the optical properties dependent on the polarization of light. Polarized light microscopy exploits this dependency and provides a sensitive tool to analyze the alignment of molecular bonds or fine structural form in man-made and natural materials.

Most biological structures exhibit some degree of anisotropy that is characteristic of their molecular architecture, such as membranes and filament arrays. A membrane is modeled as a sheet of lipid molecules in which proteins are embedded, all of which maintain some degree of orientation with respect to the plane of the membrane. Hence, tissues, cells and organelles that include extensive membranous structures, such as mitochondria and photoreceptors, can exhibit birefringence (anisotropy of refraction) and diattenuation (anisotropy of attenuation) that are characteristic of their molecular composition and architecture.

Biological structures commonly include filaments that are anisotropic by themselves, such as collagen fibrils, stress fibers made of filamentous actin and myosin, and microtubules that form the mitotic spindle. **Fig. 1** shows polarized light images of a living, dividing cell. The left image was taken with a traditional polarizing microscope and the right image was recorded with the LC-PolScope equipped for birefringence imaging. The so-called retardance image on the right was calculated based on five raw intensity images, each one not unlike the image on the left. The retardance image illustrates the clarity, sensitivity, and analytic power of quantitative imaging of anisotropic structures based on comprehensive polarization analysis that is available with the LC-PolScope.

In addition to birefringence, some biological structures also display dichroism or diattenuation in the visible light spectrum. Photoreceptors are a classic example, as the absorption of light occurs in a chromophore with a transition moment that is characteristic of a dipole and is sensitive to polarization. Because the photopigments are usually embedded in a membrane that partially orients their transition moments, the physiological response of the receptors, and often of the whole retina, is sensitive to polarization. However, while most if not all photoreceptors exhibit anisotropy of absorption, the geometric relationship between the light path in the eye and the structure of its retina can render the organ polarization insensitive, as is the case with the human eye, with the exception of a rather week sensation called "Haidinger Brush" (Le Floch et al. 2010). Yet, many animal eyes, including compound eyes of insects and eyes of fish and cephalopods, are polarization sensitive and have lead to behavioral trades that exploit this polarization sensitivity, such as navigation during an animal's migration (Reppert et al. 2010) and possibly even communication between animals of the same species (Mathger et al. 2009). By reporting a fast and comprehensive method of imaging diattenuation in biological cells and tissues, we hope to bring a new tool to studying this intriguing sense that almost completely eludes humans.

The general value of polarized light microscopy for the biomedical sciences lies in the label-free imaging of structural parameters that can be followed real-time while cells and tissues are functioning under physiological conditions. One example of this is the use of the LC-PolScope technique to image the meiotic spindle in egg cells undergoing procedures for in-vitro fertilization and for cloning (Byrne et al. 2007; Keefe et al. 2003; Liu et al. 2000).

In this article we review the use of the LC-PolScope for generating maps of birefringence measured in transparent specimens, and introduce a new use of the instrument, requiring a slight modification, to create maps of linear and circular diattenuation. We apply the instrument to measuring the birefringence and diattenuation of a nanofabricated star pattern, with some surprising results. We discuss the relation-ship between birefringence and diattenuation, how to avoid measurement artifacts, and suggest some further readings about polarized light microscopy in biology.

## 2. The LC-PolScope

Over the years, several schemes have been proposed to automate the measurement process and exploit more fully the analytic power of the polarizing microscope. These schemes invariably involve the traditional compensator, which is either rotated under computer control (Glazer et al. 1996) or replaced by electro-optical modulators, such as Pockel-cells (Allen et al. 1963), Faraday rotators (Kuhn et al. 2001), and liquid crystal variable retarders (Laude-Boulesteix et al. 2004; Oldenbourg and Mei 1995; Shribak 2011). Alternatively, one can split the imaging beam and simultaneously capture several images, analyzing each for a different polarization state (Kaminsky et al. 2007; Shribak et al. 2006). These schemes also involve quantitative intensity measurements using electronic light detectors, such as photomultipliers or charge-coupled device (CCD) cameras. For quantitative acquisition and processing algorithms that relate measured image intensities and compensator settings to optical characteristics of the specimen, see (Shribak and Oldenbourg 2003).

Here we discuss the LC-PolScope in more detail, a birefringence imaging system that was first developed at the MBL (Oldenbourg and Mei 1995; Oldenbourg and Mei 1996; Shribak and Oldenbourg 2003; Shribak and Oldenbourg 2007a; Shribak and Oldenbourg 2007b; Shribak and Oldenbourg 2008) and is commercially available from CRi Inc., now part of PerkinElmer. The LC-PolScope is equipped with a liquid crystal universal compensator that can be used to generate any polarization, from circular to elliptical to linear polarization of any orientation. Thus, the compensator works as complete polarization state generator. In this paper, we describe a new use of this device to measure the differential transmittance, also called diattenuation and dichroism, in specimens. By changing the optical setup slightly, the Birefringence LC-PolScope can be changed into the Diattenuation LC-PolScope that generates images revealing the anisotropic transmittance in man-made and natural materials at the highest resolution of the light microscope. The instructions on how to assemble the PolScope specific components on a microscope and acquire, process, and analyze polarized light images using open source software are available at OpenPolScope.org.

In the last section on "Further readings", we are referring to the many articles that discuss the design and application of the LC-PolScope as a birefringence imaging system. Here, we briefly summarize what we call the Birefringence LC-PolScope, with the aim to highlight the similarities and differences to the Diattenuation LC-PolScope that is described in this article for the first time.

### 2.1. Birefringence LC-PolScope

The optical design of the Birefringence LC-PolScope builds on the traditional polarizing microscope, introducing two essential modifications: the specimen is illuminated with nearly circularly polarized light and the traditional compensator is replaced by a liquid crystal universal compensator. In the schematic of **Fig. 2,** the universal compensator is located between the condenser lens and the light source. The universal compensator is built from two variable retarders and a linear polarizer. The variable retarders are implemented as electro-optical devices made from two liquid crystal plates. Each liquid crystal plate has a uniform retardance that depends on the voltage applied to the device. An electronic controller that is connected to a computer supplies the voltages. The polarization analyzer for circularly polarized light, placed behind the objective lens, completes the polarization optical train. A CCD or other camera, whose response is linear in the light intensity, records the specimen image that is projected by the objective lens, and specimen images are downloaded to the computer. Specialized software synchro-nizes the image acquisition process with the liquid crystal settings and implements image processing algorithms. The algorithms compute images that represent the retardance and slow axis orientation in each resolved image point.

A word on units: Retardance can be expressed as a phase angle (in degrees or radians) or as a distance (in nanometer or fraction of wavelength) between two coherent wave fronts. We use the distance between the two wavefronts that are associated with light polarized parallel to the slow and fast axes in the birefringent material. The angle of the slow axis, also called orientation or azimuth, is measured with respect to a laboratory frame of reference, which has its x-axis oriented parallel to the horizontal direction in images. The unique angles of a slow and fast axis range between 0° and 180°, with 190°, for example, equivalent to 10°.

Next, we summarize the algorithms used for acquiring and processing raw intensity images captured by the camera of the Birefringence LC-PolScope. These are algorithms for measuring parameters of linear birefringence in the absence of circular birefringence and diattenuation.

### 2.1.1. Algorithms for linear birefringence

The light intensity $I$ arriving at the camera sensor depends on several variables, including the specimen retardance ($ret$) and slow axis orientation ($azim$), the retardance of LC-A ($ret_{\text{LC-A}}$) and LC-B ($ret_{\text{LC-B}}$), the maximum amount of light $I_{\max}$ transmitted through the setup, and the minimum amount of light $I_{\min}$ transmitted when the universal compensator is adjusted for extinction:

$$I(ret, azim, ret_{\text{LC-A}}, ret_{\text{LC-B}}) = I_{\max} \; i(ret, azim, ret_{\text{LC-A}}, ret_{\text{LC-B}}) + I_{\min} \,.$$

$i(ret, azim, ret_{\text{LC-A}}, ret_{\text{LC-B}})$ represents the polarization optical train for $I_{\max} = 1$ and $I_{\min} = 0$. The complete expression for $i(ret, azim, ret_{\text{LC-A}}, ret_{\text{LC-B}})$ can be found using the Jones Calculus. A detailed discussion of the complete expression and its solutions for the specimen retardance and slow axis orientation is available in (Shribak and Oldenbourg 2003). Here we briefly present the main results.

The solutions for specimen retardance and slow axis orientation can be found, based on intensities $I_0$, $I_1$, $I_2$, $I_3$, $I_4$ representing 5 camera images. Each camera image of the specimen is recorded with a specific setting of the liquid crystal retarders of the universal compensator. The first setting generates right circularly polarized light, while the next four settings generate elliptically polarized light of equal ellipticity but with orientations of the principal axes that are rotated in steps of 45°. The five settings for $ret_{\text{LC-A}}$ and $ret_{\text{LC-B}}$ are best expressed by the differential retardance $\Delta$ by which they deviate from the setting that generates circularly polarized light. For example, circularly polarized light is generated with $ret_{\text{LC-A}} = 0.25\lambda$ and $ret_{\text{LC-B}} = 0.5\lambda$, where $\lambda$ is the wavelength of the light. Adding a "swing" value $\Delta$ to one of the retarders, e.g. $ret_{LC-A} = 0.25\lambda + \Delta$, generates elliptical polarization. Light of equal ellipticity but axis orientations that are rotated by 45° are generated by adding or subtracting $\Delta$ from $ret_{\text{LC-A}} = 0.25\lambda$ and $ret_{\text{LC-B}} = 0.5\lambda$ in turn. The following expressions represent image intensities recorded using the five settings. To reduce the complexity of the written expressions we drop all variables except for $\Delta$, such that for $ret_{\text{LC-A}} = 0.25\lambda + \Delta$ and $ret_{\text{LC-B}} = 0.5\lambda$, $I(ret, azim, ret_{\text{LC-A}}, ret_{\text{LC-B}})$ becomes $I(\Delta, 0)$:

$$I_0 = I(0,0) \;;\; I_1 = I(\Delta, 0) \;;\; I_2 = I(0, \Delta) \;;\; I_3 = I(0, -\Delta) \;;\; I_4 = I(-\Delta, 0) \,.$$

As the first setting generates circularly polarized light and leads to extinction after the light has passed through the polarization analyzer, it is called the extinction setting. The others are called elliptical settings.

Using image arithmetic, first intermediate terms $a$ and $b$ are formed

$$a = \frac{2(I_4 - I_1)}{I_1 + I_2 + I_3 + I_4 - 4I_0} \, , \ b = \frac{2(I_3 - I_2)}{I_1 + I_2 + I_3 + I_4 - 4I_0} \, ,$$

which are then used to calculate the specimen retardance $ret$ and azimuth orientation $azim$:

$$ret = \frac{\lambda}{360°} \arctan \left( \sqrt{a^2 + b^2} \, \tan(180° \frac{\Delta}{\lambda}) \right), \text{ if } I_1 + I_2 + I_3 + I_4 - 4I_0 \geq 0 \, ,$$

$$ret = \frac{\lambda}{360°} \left( 180° - \arctan \left( \sqrt{a^2 + b^2} \, \tan(180° \frac{\Delta}{\lambda}) \right) \right), \text{ if } I_1 + I_2 + I_3 + I_4 - 4I_0 < 0 \, ,$$

$$azim = \tfrac{1}{2} \arctan(b, a) \, ,$$

in which $\arctan(b, a)$ represents the $\arctan\left(\dfrac{a}{b}\right)$ with the signs of $a$ and $b$ considered independently to assign the correct quadrant for the angle.

It is interesting to note that both formulas for $ret$ and $azim$ represent ratios of differences between the raw image intensities $I_0, I_1, I_2, I_3, I_4$. Hence, the computed values of $ret$ and $azim$ are independent of the minimum intensity $I_{min}$ measured at extinction and of the maximum intensity $I_{max}$ that characterizes the throughput of the system and can vary due to shading or vignetting in the optical path.

To correct for spurious anisotropy that is contributed by optical components such as lenses, mirrors, and the liquid crystal devices, a stack of 5 images is acquired from a clear specimen area that has no birefringence. As the lenses and other optical components are still present in the optical path, the background stack records their anisotropies and can be used to remove their effect on the sample images. To do so, we first calculate background values for a and b:

$$I_{bg0} = I_{bg}(0,0) \, , \ I_{bg1} = I_{bg}(\Delta, 0) \, , \ I_{bg2} = I_{bg}(0, \Delta) \, , \ I_{bg3} = I_{bg}(0, -\Delta) \, , \ I_{bg4} = I_{bg}(-\Delta, 0) \, ;$$

$$a_{bg} = \frac{2(I_{bg4} - I_{bg1})}{I_{bg1} + I_{bg2} + I_{bg3} + I_{bg4} - 4I_{bg0}} \, , \ b_{bg} = \frac{2(I_{bg3} - I_{bg2})}{I_{bg1} + I_{bg2} + I_{bg3} + I_{bg4} - 4I_{bg0}} \, .$$

$a_{bg}$ and $b_{bg}$ are then subtracted from the specimen $a$ and $b$: $a_{corr} = a - a_{bg}$ and $b_{corr} = b - b_{bg}$; and $a_{corr}$ and $b_{corr}$ are used in the formulas for specimen retardance and azimuth, instead of $a$ and $b$. This background correction is highly effective and produces correct results for specimen retardance values that are smaller than about 30 nm. For specimen retardance values higher than about 80 nm, the background correction introduces systematic errors in the corrected values (Shribak and Oldenbourg 2003).

### 2.1.2. Example

**Fig. 3** shows retardance images recorded with the LC-PolScope, illustrating the clarity, resolution and analytic potential for examining the birefringent fine structure in living cells. It is also testimony to the opportunities still awaiting for polarized light imaging to reveal the architectural dynamics in living cells, as the structural origins of the birefringence of many of the organelles visible in **Fig. 3**, such as mitochondria and lipid droplets, remain mysterious.

The commercial LC-PolScope technique, developed and marketed by CRi Inc. (now part of Perki-nElmer), is available as an accessory to microscope stands of all major microscope manufacturers. It usually includes the universal compensator, circular polarizer, a camera with control electronics and a computer with software for image acquisition and processing. Three slightly differing versions are available, optimized for research in the life sciences (Abrio LS), industrial metrology (Abrio IM), and for in vitro fertilization and related laboratory techniques (Oosight).

When equipped for birefringence imaging, the LC-PolScope is also sensitive to diattenuation or dichroism, i.e. in the retardance image dichroic structures such as aligned pigments will be visible as anisotropic features that induce a differential phase shift (retardance) in the transmitted light. However, in the case of absorbing structures, the apparent retardance assigned by the birefringence algorithm has no direct physical meaning. Later, in the section on the "Relationship between dichroism and birefringence measurements" we will comment on this observation further.

In the next section we will describe a modification to the LC-PolScope that converts the instrument from measuring birefringence in the absence of diattenuation, to measuring only diattenuation, largely independent of differential phase shift due to birefringence.

## 2.2. Diattenuation LC-PolScope

For measuring diattenuation instead of birefringence, one only needs to remove the polarization analyzer from the optical path. **Fig. 4** shows a schematic of the Diattenuation LC-PolScope and a graph that illustrates the intensities measured on the camera sensor when illuminating a diattenuating specimen with linearly polarized light. The differential transmission, or linear diattenuation, of the specimen modulates the measured intensities as the angle of the linear polarization is rotated. The intensity shows a sinusoidal variation that has a well-established dependency on the angle of polarization, described as the generalized law of Malus (Azzam and Bashara 1973). Based on this functional dependency, it is sufficient to measure the intensity at a small number of predetermined polarization settings, as indicated in the graph. We are using four intensities to calculate parameters such as maximum and minimum transmittance, and the polarization orientation for maximum transmittance, as discussed in the section on algorithms.

To establish the transmittance of a specimen, one needs to know the incoming light intensity. Therefore, in addition to recording 4 images with the specimen in the field of view, a stack of background images must be recorded, after removing the specimen from view. In most cases, removal of the specimen is achieved by simply sliding the specimen sideways on the stage to bring a clear area into view. The need for a clear area next to the specimen of interest should be kept in mind when preparing the specimen. The recording of the background stack will also assist in removing instrument bias in the measurement of specimen diattenuation, as will be shown later.

The setup for measuring diattenuation is insensitive to birefringence, which means that the measured intensities do not depend on a differential phase shift that is likely to occur when the light passes through the sample. This feature is a direct consequence of the lack of a polarization analyzer in the imaging path. By way of explanation, any differential phase shift induced by the specimen does not affect the intensity of the light, but can affect its polarization. As the imaging path lacks a polarization analyzer, a change in polarization in the imaging light remains undetected by the setup.

To preserve the insensitivity to birefringence, it is important to avoid any optical components that may be diattenuating themselves, such as a mirror, prism, or beam splitter, in the optical path. If such compo-nents are placed downstream to the universal compensator in the light path, they lead to an instrumental bias that affects the measurement of sample diattenuation. While it is sometimes impossible to avoid components that introduce instrument bias, it is possible to measure the bias by taking recourse to the background stack introduced above. Based on the background stack, the sample images are corrected for the instrument bias, as described in the next section on algorithms. However, as always, it is best to minimize the instrument bias by choosing the right hardware und start with the cleanest raw images possible.

Next we are discussing algorithms for measuring linear diattenuation and give an example of the measured diattenuation of a Siemens star pattern etched into a thin aluminum film. This is followed by a brief summary of measuring circular diattenuation using the same setup.

## 2.2.1. Algorithm for linear diattenuation

In the following, we call linear diattenuation just diattenuation, while we will always put circular in front of diattenuation when we mean circular diattenuation.

For measuring the differential transmittance of diattenuating materials, we use the setup shown in **Fig. 4** to record images at predetermined polarization settings. As mentioned earlier, the universal compensator made of a linear polarizer followed by liquid crystal devices LC-A and LC-B is used as a universal polarizer that can generate any polarization state. For example, linearly polarized light with polarization orientation φ is generated using $ret_{LC-B}$=0.25 λ and $ret_{LC-A}$=λ·φ/180°. Let us consider the intensity of light that has passed through first the universal polarizer, followed by a uniformly diattenuating specimen, and is then projected onto the camera. As we change the orientation $\phi$ of the linear polarization from 0° (horizontal), to 90° (vertical), to 180° (horizontal again), the recorded intensity $I(\phi)$ varies according to the following expression (generalized law of Malus):

$$I(\phi) = \tfrac{1}{2}\left(I_{max} + I_{min} + \left(I_{max} - I_{min}\right)\cos\left(2(\phi - azimuth)\right)\right) = \tfrac{1}{2}I_i\left(T_{max} + T_{min} + \left(T_{max} - T_{min}\right)\cos\left(2(\phi - azimuth)\right)\right),$$

$I_i$ is the initial intensity measured without diattenuating specimen,

$I_{max} = T_{max} \cdot I_i$ is maximum intensity measured with diattenuating specimen, $T_{max}$ is its maximum transmittance,

$I_{min} = T_{min} \cdot I_i$ is minimum intensity measured with diattenuating specimen, $T_{min}$ is its minimum transmittance,

$azimuth$ is the polarization orientation that results in maximum transmittance.

To comprehensively describe the characteristics of the specimen's transmittance, we determine the $azimuth$, its mean transmittance $T_{mean}$, and its *diattenuation* (see for example (Chipman 2010)):

$$T_{mean} = \tfrac{1}{2}\left(T_{max} + T_{min}\right), \ diattenuation = \frac{T_{max} - T_{min}}{T_{max} + T_{min}}$$

To determine these characteristics, we measure the intensity for four linear polarization states whose orientations are rotated by 45° to each other in a laboratory frame of reference:

$$I_0 = I(0°), \ I_{45} = I(45°), \ I_{90} = I(90°), \ I_{135} = I(135°)$$

By measuring intensities $I_0$, $I_{45}$, $I_{90}$, $I_{135}$ and entering them together with the corresponding polarization orientations into the expression for $I(\phi)$, we have 4 equations that can be solved for the unknowns *azimuth*, $T_{mean}$, and *diattenuation*. We first form intermediate results $a$, $b$, and $c$:

$$a = I_0 - I_{90} \, , \; b = I_{45} - I_{135} \, , \; c = I_0 + I_{45} + I_{90} + I_{135} \, ,$$

and then express the unknowns as a function of those intermediate results:

$$azim = \frac{\arctan(b, a)}{2} \, ,$$

$$diattenuation = \frac{I_{\max} - I_{\min}}{I_{\max} + I_{\min}} = \frac{T_{\max} - T_{\min}}{T_{\max} + T_{\min}} = \frac{2\sqrt{a^2 + b^2}}{c} \, , \; \text{values } 0 \ldots 1.$$

In addition to diattenuation, the differential transmission may also be expressed in terms of the ratio of $I_{\max}$ to $I_{\min}$:

$$ratio = \frac{I_{\max}}{I_{\min}} = \frac{T_{\max}}{T_{\min}} = \frac{1 + diattenuation}{1 - diattenuation} \, , \; \text{values } 1 \ldots \text{infinity}$$

To calculate the mean transmittance, we make use of the background stack of images and recognize $I_i$ as the mean background intensity:

$$I_i = \frac{I_{bg0} + I_{bg45} + I_{bg90} + I_{bg135}}{4} \, , \; \text{hence}$$

$$T_{\text{mean}} = \frac{I_0 + I_{45} + I_{90} + I_{135}}{I_{bg0} + I_{bg45} + I_{bg90} + I_{bg135}} \, .$$

To account for spurious differential attenuation caused by microscope optics, including lenses, mirrors, filters, and the liquid crystal devices, the background stack is also used to correct the sample image stack. To that end, each sample intensity is divided by the corresponding background transmittance, such as $I_0$ is divided by $T_{bg0} = \dfrac{I_{bg0}}{I_i}$, hence:

$$I_{corr0} = I_0 \frac{I_i}{I_{bg0}} \, , \; I_{corr45} = I_{45} \frac{I_i}{I_{bg45}} \, , \; I_{corr90} = I_{90} \frac{I_i}{I_{bg90}} \, , \; \text{and } I_{corr135} = I_{135} \frac{I_i}{I_{bg135}}$$

The corrected values $I_{corr0}$ through $I_{corr135}$ are then used in the expressions for the intermediate results $a$, $b$, and $c$.

As mentioned earlier, all expressions are implemented as image arithmetic operations and are applied to all pixels simultaneously, resulting in *azimuth* and *diattenuation* images that represent quantitative maps of differential transmittance, polarization orientation for maximum transmittance, and mean transmittance, all measured at the resolution of the microscope optics.

An easy way to recognize the efficacy of correcting the instrumental bias is to record a background stack, followed by a sample stack, both stacks imaging the same clear area. Processing the sample stack without background correction is likely to result in low diattenuation varying slowly across the field of view. The finite diattenuation in this case is caused by spurious dichroism in components of the microscope optical path. The changes in intensity incurred by spurious polarization dependent transmission are carried through the optical system, are projected into the image plane, and are measured as variations in the background diattenuation. However, the data recorded in the background stack record the variations and can be used to remove their effect on the sample data. Hence, a sample stack of a clear area that is corrected in the above manner and then processed, will show minimal and uniform diattenuation across the image plane. Usually the minimal diattenuation is not zero, but varies randomly from pixel to pixel and is caused by the noise (typically shot noise) in the collected image data. These random variations in diattenuation due to shot noise set a lower limit on the sample diattenuation that can be reliably measured.

When inspecting the azimuth data that result from processing the sample stack without background correction, one typically recognizes azimuth values that vary slowly across the image. Processing the same sample data with background correction results in an azimuth image that is uniform across the image and azimuth data randomly vary from pixel to pixel between 0° and 180°. This random variation is again caused by the shot noise in the raw image data.

Next, we illustrate the sensitivity and resolution afforded by the Diattenuation LC-PolScope, demonstrated by a nanofabricated star pattern etched into a thin metal film.

## 2.2.2. Example

**Fig. 5** shows the mean transmittance, diattenuation, and apparent retardance of a Siemens star etched by electron lithography into a thin metal film (Oldenbourg et al. 1996). The radial spoke pattern of 36 wedge pairs represents a grating of varying orientation and with a periodicity that increases with increasing distance from the center of the pattern. The image was projected by a 40x Fluor (Nikon) oil immersion objective with a diaphragm in the back focal plane to set the objective's numerical aperture to 0.5. The average transmittance near the central part of **Fig. 5 B** displays the expected resolution according to the wavelength $\lambda = 630$ nm and the *NA* of 0.5 that was set for both, objective and condenser lens. In addition, the anisotropy of the pattern, which is revealed in panels **C** and **D** by the polarization dependent phase shift (apparent retardance) and transmittance (diattenuation) shows interesting features, which we describe here qualitatively, but leave a quantitative evaluation to a future publication.

The modulation of the intensity approaches zero for a radius of 3.9 μm, which corresponds to a periodicity of $2 \cdot 3.9 \cdot \pi / 36 = 0.68 \, \mu m$, close to the expected value of the resolution limit for the *NA* and wavelength used ($\lambda / (2 \, NA) = 0.63 \, \mu m$). The modulation increases for larger radii, while for smaller radii the modulation remains flat.

At an even smaller radius of 1.8 μm, the transmittance increases almost twofold to remain high until the finer and finer grating ends at the central disk (radius 0.6 μm). The grating at a radius of 1.8 μm has a periodicity of 0.315 μm, which is half the wavelength of the light used.

Panel **C** in **Fig. 5** shows the retardance near the center and panel **D** shows its differential transmittance for linear polarization, measured with the diattenuation mode of the LC-PolScope. For radii larger than 3.9 μm, where wedge edges are resolved, low diattenuation is observed near the wedge edges. However, in the unresolved central part of the pattern, one can recognize two bright, highly diattenuating rings separated by a dark ring. The outer ring is the most diattenuating feature of the pattern, but transmits low light. The inner ring transmits more light but is less diattenuating.

The periodicity of the pattern that is responsible for the innermost ring of anisotropy is smaller than half the wavelength of light. Grid patterns of those dimensions are called wire grid polarizers and their action is well understood (Bennett 2010). Wire grid polarizers reflect the light whose electric field is polarized parallel to the linear wires. Hence, maximum transmittance through the wire grid is observed for the polarization that is perpendicular to the wire orientation, which corresponds to the azimuthal alignment of the lines that indicate the orientation of high transmittance in the innermost ring of panel **D** in **Fig. 5**.

The origin of the high diattenuation and phase shift observed for periodicities between $\lambda/2$ and $\lambda$ of the wavelength must be related to a resonance in the interaction between the wire grid pattern and the light field of similar periodicities. We are leaving a quantitative evaluation of those phenomena to a future publication, but point out that the retardance measurement using the Birefringence LC-PolScope includes both effects, differential transmittance and differential phase shift. Hence, the apparent retardance cannot be interpreted as a phase shift alone, but its interpretation has to also account for the diattenuation, which is measured by the Diattenuation LC-PolScope without interference from differential phase shifts, as discussed in section 3.

### 2.2.3. Algorithm for circular diattenuation

For measuring circular diattenuation, we can use the same setup as shown in **Fig. 4** and acquire images of the specimen when it is trans-illuminated with left and right circularly polarized light, instead of linearly polarized light. The two circular polarizations can be generated using $\text{ret}_{\text{LC-A}} = 0.25\lambda$ and $\text{ret}_{\text{LC-A}} = 0.75\lambda$, keeping $\text{ret}_{\text{LC-B}}$ constant at $0.5\lambda$. Specimen components that exhibit circular diattenuation will have a non-zero differential transmittance $T_{left} - T_{right}$, with $T_{left}$ and $T_{right}$ the transmittance for left and right circularly polarized light, respectively. Circular diattenuation has no azimuth orientation associated with it.

Similar to the transmittance for linearly polarized light, we define the material parameters for circularly polarized light as mean circular transmittance $CT_{\text{mean}}$ and *circular_diattenuation*:

$$CT_{\text{mean}} = \frac{1}{2}\left(T_{\text{left}} + T_{\text{right}}\right), \text{ and}$$

$$circular\_diattenuation = \frac{T_{\text{left}} - T_{right}}{T_{\text{left}} + T_{right}},$$

with $I_{\text{left}} = T_{\text{left}} \cdot I_i$ and $I_{\text{right}} = T_{\text{right}} \cdot I_i$ and $I_i$ the incoming intensity. Again, we find $I_i$ using the intensities measured in a clear area of the specimen, $I_i = \frac{1}{2}\left(I_{\text{bg left}} + I_{\text{bg right}}\right)$, and use $I_{\text{bg left}}$ and $I_{\text{bg right}}$ to correct for possible instrument bias:

$$I_{\text{corr left}} = I_{\text{left}} \frac{I_i}{I_{\text{bg left}}}, \ I_{\text{corr right}} = I_{\text{right}} \frac{I_i}{I_{\text{bg right}}},$$

$$CT_{\text{mean}} = \frac{I_{\text{corr left}} + I_{\text{corr right}}}{2\,I_i}, \text{ and}$$

$$circular\ diattenuation = \frac{I_{\text{corr left}} - I_{\text{corr right}}}{I_{\text{corr left}} + I_{\text{corr right}}}.$$

## *3. Relationship between diattenuation and birefringence measurements*

### 3.1. Birefringence and orientation of slow axis

Birefringence occurs when there is a molecular order, when the average molecular orientation is non-random, as in crystals or in aligned polymeric materials. For a given propagation direction of light, molecular order usually gives the material two orthogonal optical axes, with the index of refraction for light polarized along one axis being different from the other axis. The difference between the two indices of refraction is called birefringence, and is an intrinsic property of the material:

$$\textbf{birefringence} = \Delta n = n_{\parallel} - n_{\perp},$$

where $n_{\parallel}$ and $n_{\perp}$ are the refractive indices for light polarized parallel and perpendicular to the two optical axes. We note for the above expression, birefringence refers to the cross section between the refractive index ellipsoid of the material and the plane whose normal is the propagation direction of the light. This is also called the birefringence of the section, in contrast to the maximum or substance birefringence observed for a uniaxial material when the optic axis is located in the plane of observation (Hartshorne and Stuart 1960).

Light polarized parallel to one axis travels at a different speed through the sample than light polarized perpendicular to that axis. As a result, the two light components, which were in phase when they entered the sample, are retarded at a different rate and exit the sample out of phase. Using the polarizing microscope, one can measure this differential retardation, also called retardance, and thereby quantify the magnitude and orientation of molecular order in the specimen. Retardance is an extrinsic property of the material and a product of the birefringence and the path length $l$ through the material:

$$\textbf{retardance} = R = \Delta n \cdot l.$$

Retardance refers to the differential phase shift between orthogonal polarization components. However, if differential transmittance cannot be neglected in the specimen, the Birefringence LC-PolScope measures a retardance that not only reports the differential phase shift but is also affected by the differential transmittance in the material (see section "Measuring differential transmittance as retardance").

Birefringence also has an orientation associated with it. The orientation refers to the specimen's optical axes of which one is called the **fast axis** and the other the **slow axis**. Light polarized parallel to the slow axis experiences a high refractive index and travels slower than light polarized parallel to the fast axis. In materials that are built from aligned filamentous molecules, the slow axis is typically parallel to the average orientation of the filaments. Birefringence orientation always correlates with molecular orientation, which usually changes from point to point in the specimen.

Like retardance, the orientation of the slow axis measured by the Birefringence LC-PolScope is also affected by diattenuation present in the material, as discussed later.

For a more comprehensive discussion of birefringence of optically anisotropic materials, see (Hartshorne and Stuart 1960; Hecht 2002; Shurcliff 1962).

### 3.2. Linear diattenuation and orientation of maximum transmittance

Linear diattenuation is a material property that can occur in absorbing materials in which the light absorbing molecules or fine structure are arranged with a preferred orientation. Diattenuation refers to the difference in transmittance for light polarized at specific angles to the principal axis of alignment.

Diattenuation and dichroism describe similar material properties and their relationship can be confusing. Additional confusion arises from the fact that dichroism has two related but distinct meanings in optics. A dichroic material is either one which causes visible light to be split up into distinct beams of different wavelengths, or one in which light rays having different polarizations are absorbed at different rates. The latter meaning describes a similar material property as diattenuation. Dichroism describes the property in terms of the material's absorption and diattenuation in terms of its transmittance. As we are concerned with light that is transmitted through the sample, and not the part that is absorbed, we refer to the instrument as Diattenuation LC-PolScope.

For a given propagation direction of light, the molecular alignment creates two principal axes in the material. When light is polarized parallel to one axis, the material has maximum transmittance $T_{max}$, and when it is polarized parallel to the other axis, its transmittance $T_{min}$ is a minimum. The two axes are necessarily orthogonal to each other. When light of intensity $I_i$ is polarized at an arbitrary angle $\theta$ and enters the material, it is split into two components; one ($I_i \cos^2(\theta)$) polarized parallel to the direction of high transmittance and one ($I_i \sin^2(\theta)$) is polarized perpendicular to it. When the two components leave the material boundaries, they recombine and are detected as light that was transmitted with a transmittance $T(\theta)$ that depends on the polarization angle $\theta$ of the incoming light. The total detected intensity is simply the sum of the intensity of both polarized components in the absence of a polarizing element between the material and the detector. Thus,

$$I_i T(\theta) = I_i T_{max} \cos^2 \theta + I_i T_{min} \sin^2 \theta = \tfrac{1}{2} I_i \left( T_{max} + T_{min} + \left( T_{max} - T_{min} \right) \cos 2\theta \right)$$

The above expression is identical to the earlier expression that was identified as a generalization of the law of Malus, with the *azimuth* variable set to zero. When *azimuth* is zero, the principal axes of the material coincide with the laboratory frame of reference.

With the Diattenuation LC-PolScope we can measure the mean transmittance $T_{mean} = \tfrac{1}{2} \left( T_{max} + T_{min} \right)$, the differential transmittance, and the angle of maximum transmittance in the form of high-resolution images of the sample, documenting even subtle variations that reflect the molecular order changing from point to point in the sample. Importantly, diattenuation measurements are largely unaffected by differential phase shifts that are induced by concomitant birefringence in the sample. However, as mentioned earlier, the Birefringence LC-PolScope is sensitive to both, birefringence and diattenuation. Next, we briefly discuss this somewhat confusing distinction between the two modes of measurement.

### 3.3. Measuring diattenuation as retardance

As mentioned earlier, the Birefringence LC-PolScope is sensitive to both, birefringence and diattenuation. The polarization analyzer in the light path of the Birefringence LC-PolScope makes the instrument sensitive to changes in polarization, regardless if they are incurred by birefringence or diattenuation in the specimen. By way of explanation, let's consider a diattenuating structure like the wire grid polarizer formed by the unresolved pattern near the center of the Siemens star in **Fig. 5**. When light of arbitrary polarization angle enters the grid, it is split into two components, one polarized parallel and the other perpendicular to the grid lines, and each is attenuated at a different rate. When the components leave the grid, their superposition will generally lead to a polarization that is different from the light entering the grid, because the amplitudes of the two components are different from before the grid. As a change in polarization is the primary signal that is analyzed by the birefringence measurement setup of **Fig. 2**, the grid will cause an apparent retardance solely based on the fact that the two polarization components were subject to differential transmittance. However, if referred to as a differential phase shift, the diattenuation-induced retardance has no physical meaning and has to be considered a measurement artifact. In addition, diattenuation-induced retardance has an apparent slow axis orientation that is rotated by 45° with respect to the axis of maximum transmittance. (One can prove this point using the Jones Calculus, for example.) The apparent retardance and slow axis orientation is added to any real differential phase shift that is likely to occur for a diattenuating material. The superposition of apparent and real phase shifts that are measured together in the Birefringence LC-PolScope lead to the skewed angle of slow axis orientation observed in the retardance image of the central part of the Siemens star in **Fig. 5C**.

This latter feature can actually help in identifying a retardance measurement that is "contaminated" by differential transmittance. Let us look again at the slow axis orientation in **Fig. 5C**. It has an angle that is neither parallel nor perpendicular to the grid lines and is therefore suspect. There is no physical reason for the true differential phase shift to have a symmetry, i.e. orientation of its slow and fast axis, that is different from the symmetry of the grating itself. The twist of the slow axis in **Fig. 5C** is caused by both, the grid and the measurement apparatus together, and is not a physical property of the grid alone. However, as the measurement of the differential transmittance is unaffected by differential phase shift, we believe the two measurements together can be used to remove the effect of differential transmission on the measured retardance. We leave a more detailed analysis of this statement to a future publication.

In the last section, we briefly touch on the fact that diattenuation is always accompanied by differential phase shift, making the feature of insensitivity to phase shift especially valuable in the Diattenuation LC-PolScope.

## 3.4. Physical relationship between birefringence and diattenuation

Diattenuation and birefringence refer to the polarization dependence of the two, more basic phenomena of absorption and refraction, respectively. Both these phenomena are the consequence of light interacting with matter and are therefore closely related, to the point that one would not exist without the other (Feynman et al. 1963). We briefly discuss this close relationship, with the intent to develop an intuitive understanding for the reader.

The refractive index is an expression of refraction that occurs because of the electric field emitted by electrons that oscillate in a material in response to the incident light. The oscillations of electrons that are forced by the incident field cause secondary radiation that has a fixed phase relationship with the incoming light. The secondary radiation, when combined coherently with the incident radiation, results in a slowed wave front, which leads to the phenomenon of refraction. The refractive index directly describes the degree of slowdown. The amplitude and phase response of the electron cloud depends on the wavelength (or frequency) of the light – a phenomenon called dispersion. At frequencies near a resonance, the secondary radiation can be opposite in phase to the incident radiation, reducing the amplitude of the transmitted radiation significantly and leading to the phenomenon of absorption or reduced transmittance. Birefringence and diattenuation, expressing the polarization dependence of refraction and transmittance, occur when the oscillating electron clouds have restoring forces that depend on the mutual angle between the polarization of the incident light and the orientation of molecular and/or fine structural order of the material.

Most biological materials, like carbohydrates, proteins and nucleic acids, absorb in the ultraviolet at frequencies that are higher than the visible spectrum of light. The material refractive index in the visible regime is an expression of the strength and anisotropy of the resonances that occur in the ultraviolet. Usually, the refractive index and birefringence of those materials vary little over a large frequency range in which no absorption occurs.

Other biological materials, like pigments and light absorbing proteins in the retina, absorb in the visible spectrum, resulting in absorption spectra that are quite specific to the pigments involved. Therefore, spectrally resolved absorption and diattenuation measurements have the potential to provide specificity to molecular order and molecular composition.

## 4. Further readings

The classic books on polarizing microscopy were published several decades ago, when optical phase microscopy was in its heydays (Hartshorne and Stuart 1960; Hartshorne and Stuart 1964). Several decades before Hartshorne, W. J. Schmidt published 2 celebrated monographs that to this day report the most comprehensive surveys of biological specimens observed under polarized light (Schmidt 1924; Schmidt 1937). Inspired by Schmidt's observations, Shinya Inoué significantly improved the sensitivity of the instrument and made seminal contributions to our understanding of the architectural dynamics inside living cells (Inoue et al. 1975; Inoué and Hyde 1957; Inoue and Sato 1967). For many lucid discussions of polarized light microscopy and its application to biology, we refer the reader to the many articles and books by Inoué, including his recently published Collected Works (Inoue 2002; Inoué 1986; Inoué 2003; Inoué 2008; Inoue and Oldenbourg 1998). Tinoco and collaborators reviewed the imaging of dichroism and diattenuation in biological materials (Tinoco et al. 1987). Jin et al. applied the rotating-polarizer method to studying the linear birefringence and diattenuation in cerebral amyloid pathologies (Jin et al. 2003). The following articles include the LC-PolScope in their discussion of various aspects of polarizing microscopy (Oldenbourg 1999; Oldenbourg 2005; Oldenbourg 2007; Oldenbourg and Shribak 2010; Shribak and Oldenbourg 2003). A glossary of polarization optical terms, such as retardance and slow axis, is contained in (Oldenbourg 2013)

## 5. Conclusion

Polarized light microscopy allows one to nondestructively follow the dynamic organization of living cells and tissues at the microscopic as well as submicroscopic levels. Imaging with polarized light reveals information about the organization of the endogenous molecules that built the complex and highly dynamic architecture of cells and tissues. We have shown that the LC-PolScope can be used for imaging birefringence, linear and circular diattenuation, and is thus a versatile tool to reveal information about structural parameters, such as the alignment of molecular bonds and their transition moments, and of the chirality of molecules.

We believe, with current instrumentation we have only scratched the surface of the potential of polarized light microscopy and we look forward to more advances that will emerge with combining hardware and software components with new insights into the interaction of polarized light with the building blocks of man-made and natural materials.

## 6. Acknowledgments

The authors gratefully acknowledge many years of illuminating discussions on polarized light microscopy with Shinya Inoué of the MBL. Grant Harris and Amitabh Verma implemented image acquisition and processing software.

This work was supported by a fellowship of the Human Frontiers Science Program awarded to SM and a grant from the National Institute of Biomedical Imaging and Bioengineering (grant Nr. R01EB002045) awarded to RO. MS acknowledges support from the National Institute of General Medical Sciences/NIH (grant R01-GM101701).

## 8. Figures

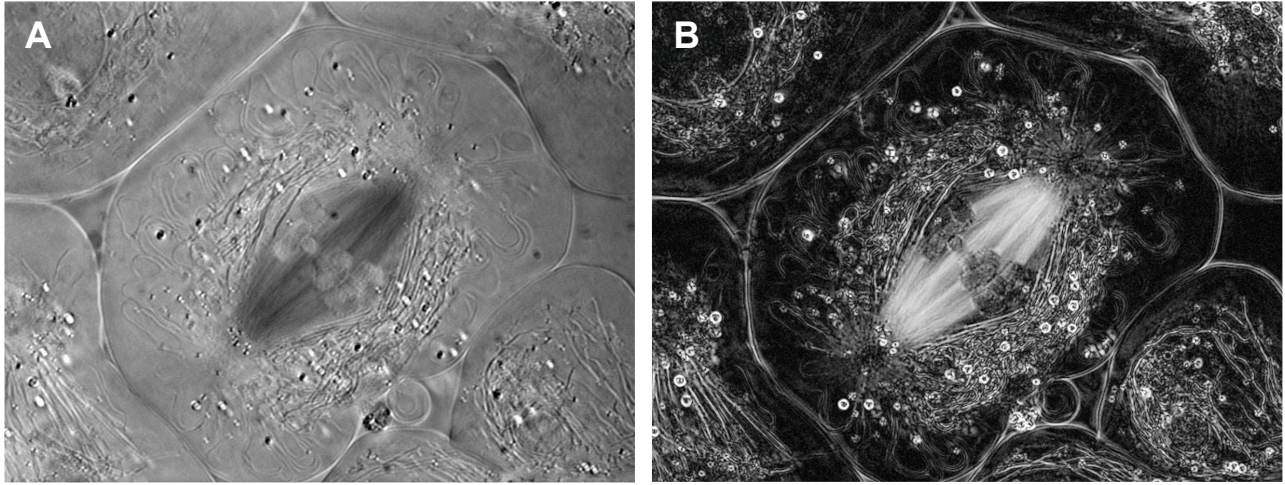

**Figure 1** Polarized light images of a primary spermatocyte from the crane fly, *Nephrotoma suturalis*, taken (**A**) with the traditional polarized light microscope and (**B**) with the LC-PolScope. In both images variation of brightness is caused by the birefringence of cell structures, like the meiotic spindle near the center. However, in the traditional polarized light image, brightness is a function of both, the orientation of the spindle with respect to the polarizers and compensators of the instrument, and a function of the magnitude of birefringence. In the LC-PolScope image, brightness is directly proportional to the birefringence of the specimen, independent of the orientation of the birefringence axis (slow axis). The LC-PolScope image of the flattened cell reveals with great clarity the birefringence of spindle microtubules extending from the chromosomes to the spindle poles, and of astral microtubules radiating from the centrosomes located near the poles. The birefringence of other cell organelles such as elongated mitochondria, surrounding the spindle like a mantle, and small spherical lipid droplets, is also evident against the dark background of the cytoplasm. The pole-to-pole distance is approximately 25μm.

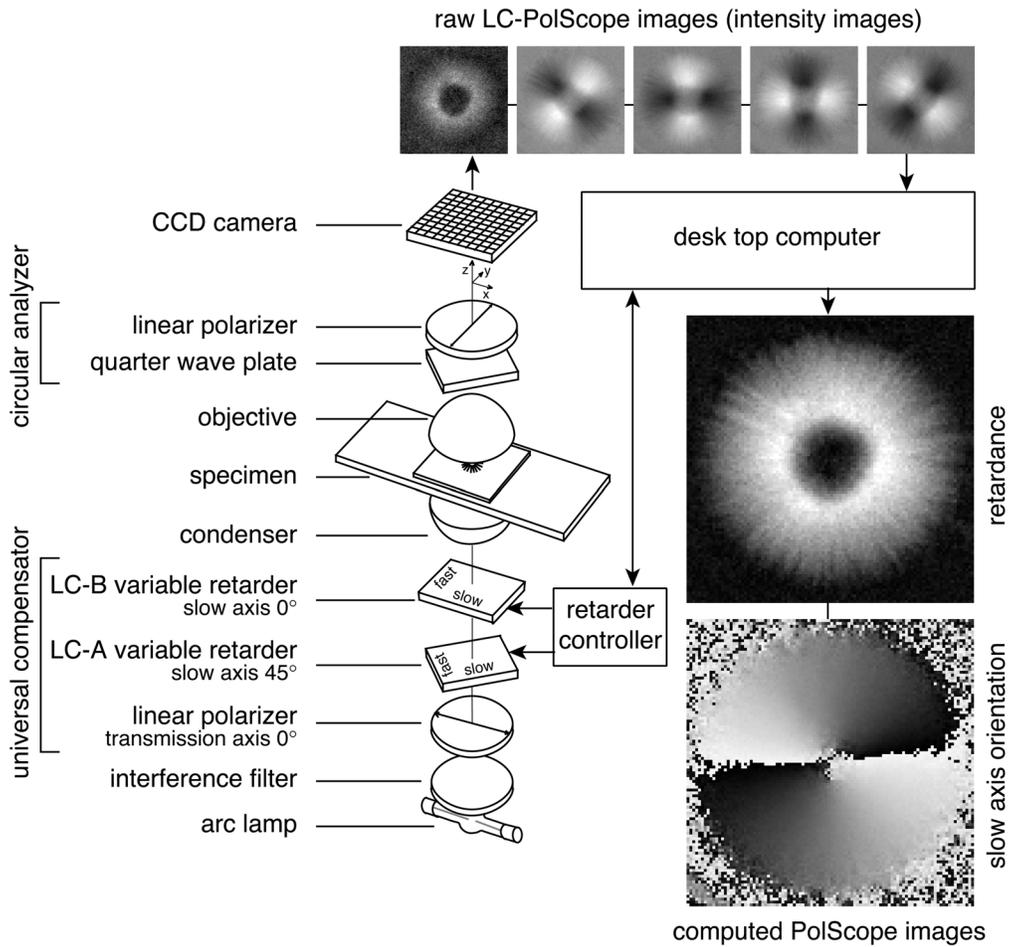

**Figure 2** Schematic of the LC-PolScope for birefringence imaging. The optical design (left) builds on the traditional polarized light microscope with the conventional compensator replaced by two variable retarders LC-A and LC-B. The polarization analyzer passes circularly polarized light and is typically built from a linear polarizer and a quarter wave plate. Images of the specimen (top row, aster isolated from surf clam egg) are captured at five predetermined retarder settings, which cause the specimen to be illuminated with circularly polarized light (1st, left most image) and with elliptically polarized light of different axis orientations (2nd to 5th image). Based on the raw PolScope images, the computer calculates the retardance image and the slow axis orientation or azimuth image using specific algorithm. The LC-PolScope requires the use of narrow bandwidth (≤40 nm) or monochromatic light.

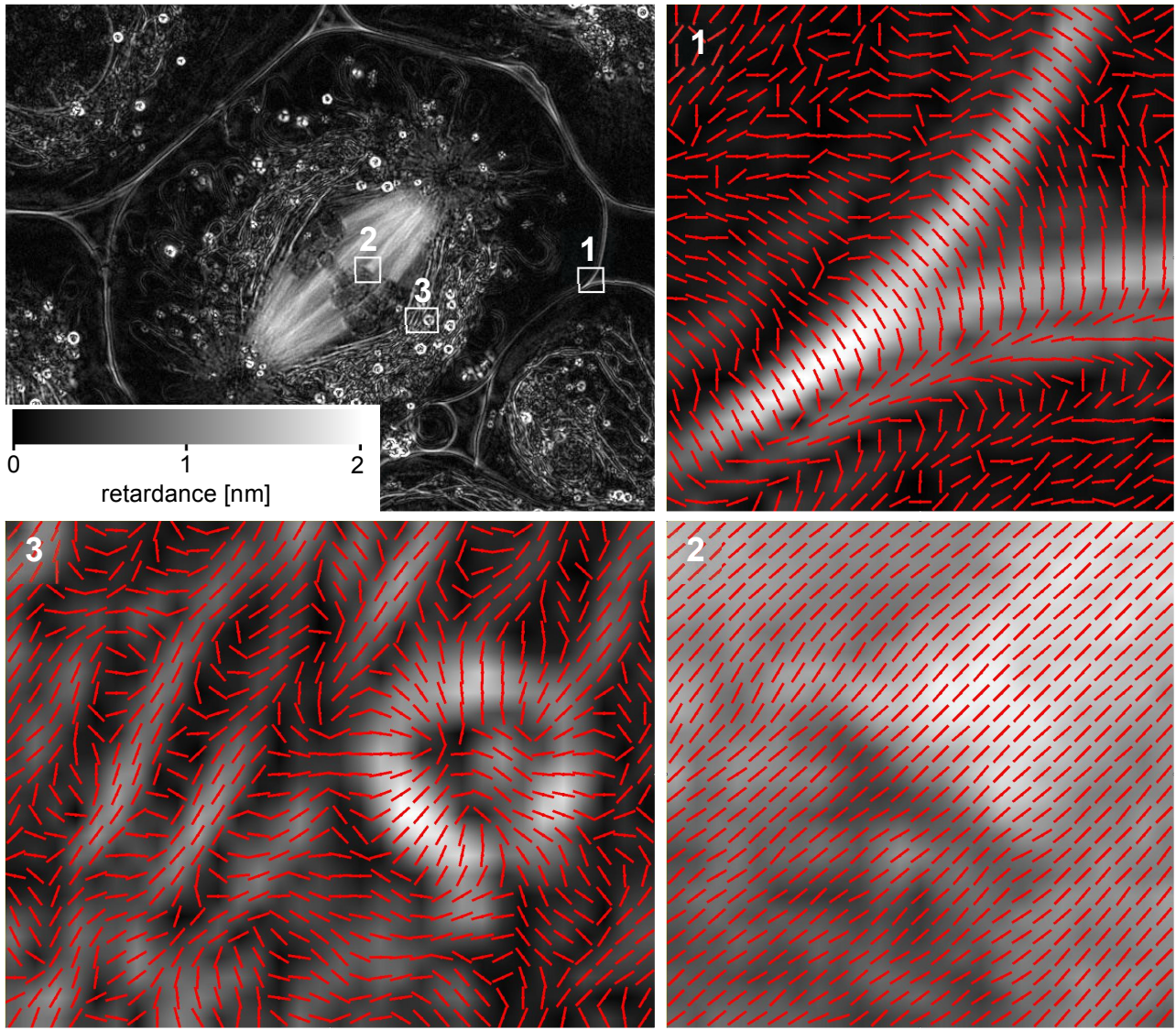

**Figure 3** Birefringence LC-PolScope images with slow axis lines of the primary spermatocyte shown in Fig. 1. At the top left, the retardance image of the whole cell is dominated by the meiotic spindle extending between two poles and with chromosomes aligned at the metaphase plate. Regions 1, 2 and 3 are shown enlarged with red lines indicating the slow axis direction for each pixel. The underlying retardance images were enlarged using bilinear interpolation. **Region 1** shows a cross section through the cell membrane in which the slow axis is oriented perpendicular to the plane of the membrane. **Region 2** identifies a kinetochore with centromere and attached kinetochore fiber, with slow axes parallel to the microtubule bundles. **Region 3** shows on the right a lipid droplet with a highly birefringent shell and slow axes perpendicular to the droplet's surface. The left side of Region 3 shows parts of mitochondria that surround the meiotic spindle like a mantle of birefringent tubes, with a slow axis parallel to the tube axis. (The cell was prepared and the image recorded by James R. LaFountain, University at Buffalo)

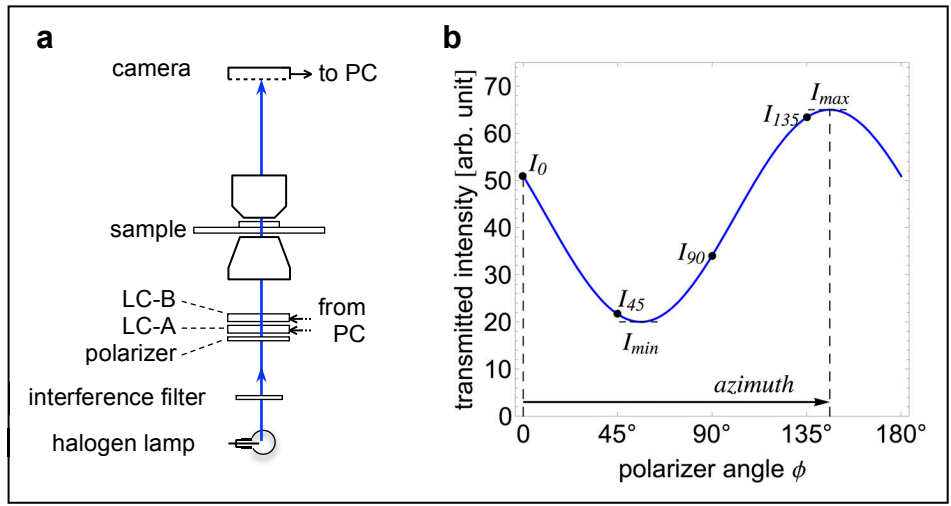

**Figure 4** Diattenuation measurement scheme. The schematic of the optical setup includes the universal compensator (polarizer, LC-A, LC-B) that is used to create linearly polarized light for illuminating the sample. No polarization analyzer is present in the imaging path. (The reverse is possible too, i.e. using unpolarized light as illumination and analyzing the linear polarization of the imaging light.) The graph on the right shows the expected intensity measured in a region of uniform dichroism with a maximum transmittance for light that is linearly polarized at an azimuth orientation of 146°. The solid line is the expected sinusoidal variation when continuously varying the polarization orientation, with intensities labeled for orientations 0° (horizontal), 45°, 90°, and 135°.

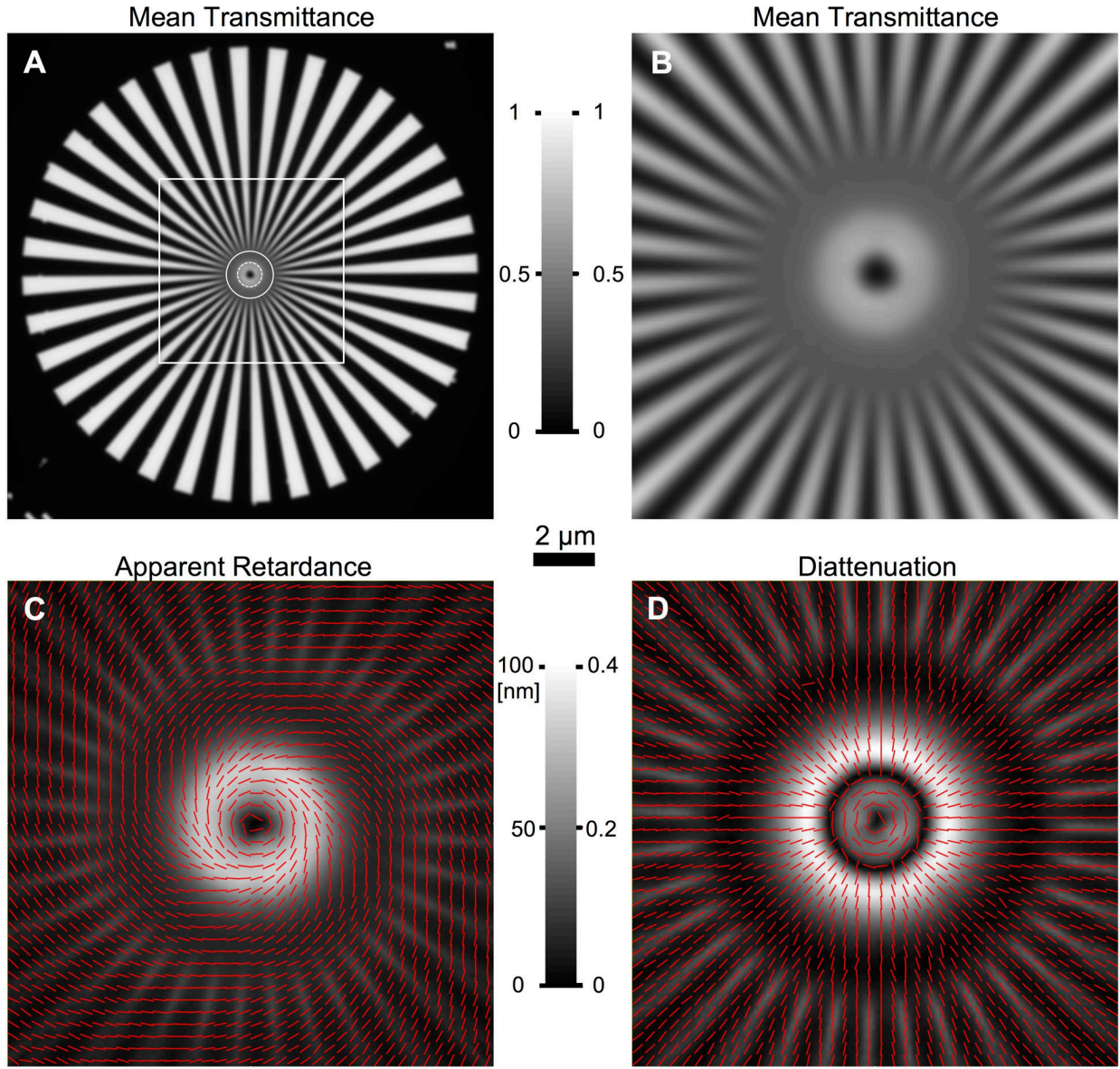

**Figure 5** LC-PolScope images of a Siemens star etched into a 50 nm thick aluminum film on cover glass using electron lithography. The cover glass with film was mounted on a microscope slide using Permount. **A** Average transmittance of the star pattern, which consists of 36 wedge pairs and has an outside diameter of 75 μm. Near the outer edge along the circumference the grating period is 6.5 μm, which decreases continuously towards the smallest period of 0.1 μm near the inner black disk (dia. 1.2 μm). Near the white dashed circle, the period is 0.5 λ, and λ near the solid circle, with λ the wavelength of the illuminating light (630 nm). The square outlines the magnified portion in panels B to D. **B to D** Central portion of the Siemens star imaged with the LC-PolScope in different imaging modes: **B** Mean transmittance. Gray scale legend between panels A and B indicates mean transmittance and applies to both panels. **C** Apparent retardance of the central pattern with red lines indicating the slow axis orientation. **D** Diattenuation of the central pattern with red lines indicating the orientation of the maximum transmittance. In panels C and D lines are drawn for every 3rd pixel of the original image. Gray scale legend between panels C and D shows retardance scale for left and anisotropy scale for right panel. Scale bar in central part of the Figure applies for images in panels B to D.